\documentclass[a4paper,twocolumn]{esapub2005} 
\usepackage{times}
\usepackage{natbib}
\usepackage{epsfig}
\usepackage{graphicx}
\title{The spectral resolving power of irregularly sampled time series}
\author{Frank P. Pijpers}
\affil{Space and Atmospheric Physics Group, Imperial College London, 
Blackett lab., Prince Consort Road, London SW7 2BW. 
e-mail: F.Pijpers@imperial.ac.uk}

\begin{document}


\maketitle

\begin{abstract}
A method is presented for investigating the periodic signal content of time
series in which a number of signals is present, such as arising from the
observation of multiperiodic oscillating stars in observational
asteroseismology.
Standard Fourier analysis tends only to be
effective in cases when the data are perfectly regularly
sampled. During normal telescope operation it is often the case that
there are large, diurnal, gaps in the data, that data are missing, or
that the data are not regularly sampled at all. For this reason it is
advantageous to perform the analysis as much as possible in the time
domain. Furthermore, for quantitative analyses of the frequency
content and power of all real signals, it is of importance to have
good estimates of the errors on these parameters. This is easiest to
perform if one can use linear combinations of the measurements. Here
such a linear method is described. The method is based in part on well-known
techniques in radio technology used in every FM radio receiver, and in
part on the SOLA inverse method.
\end{abstract}

\section{Introduction}
Low-amplitude solar and stellar oscillations can be considered as a 
linear superposition of normal modes of resonance. In theory one
indexes each by a radial order $n$ for the mode, and the degree
$\ell$, and order $m$ of the spherical harmonic functions which
describe the azimuthal spatial behaviour of the normal
modes.
For solar-like oscillations, excited stochastically by turbulent
convection, there are usually many oscillation modes present at any
epoch, with comparable amplitudes. If a time series, in which there are 
a number of periodic signals present, is sampled perfectly regularly 
the problem of determining the frequencies can be solved in a
straightforward manner using a Fourier transform and there is a single
peak for each real frequency present in the data. 

If there are gaps in 
the data or if the data is irregularly sampled there will be more than 
one peak in the Fourier spectrum for each signal frequency in the data. The
shape of this response for each signal frequency that is present in the data
is known as the `window function'. Clearly, if there are a large number of
real signal frequencies in gapped or irregularly sampled data, the Fourier
transform very quickly becomes very complicated and hence it becomes
very difficult to determine which are true frequencies, and which are
aliases generated by the sampling.

There exist a number of techniques for resolving the problem of
determining amplitudes, frequencies and phases of periodic signals in
irregularly sampled, gapped data. Some attempt to `fill gaps' using
autoregressive methods (cf. \cite{FaUl82},
\cite{BJCD90}, \cite{Foea99}). Other methods
attempt to do a least-squares fitting of sinusoids in the time domain
(cf. \cite{Sc82}, \cite{Sc89},
\cite{Ko99}) or to define a best-fit spectrum in a
least-squares sense, but mollified to take into account that this
problem is ill-posed since it is a form of an inverse problem
(\cite{SLL00}, \cite{VSW00}).

Apart from the problems of sampling, there is also the problem of noise
in the data.
In principle a star can produce `intrinsic noise', through the stellar
equivalent of granulation. The telescope and instrumentation can also 
introduce noise into time-series which in an ideal case would amount
to the photon noise due to finite integration times. Such noise might
be white, i.e. independent of frequency, but in some cases it could be
confined to bands of frequencies. It therefore makes some sense to
`chop up' the full frequency  range available, set by the sampling and
length of the time series, into  different regimes. 
\cite{ORPPN} refer to this as a `multiresolution approach'.

\section {The method}
In the approach of \cite{ORPPN} the series is convolved with every
member of a set of Gaussians, with successively increasing 
widths $\Delta_{j+1}=2\Delta_j$ and the differences between these 
successive convolutions are computed. If the data were regularly sampled each
of these differences would correspond in the frequency domain to
having applied a band-pass filter to the data. If all the different
multiresolution levels were to be co-added the original data would be
reproduced. The convolution is carried out in the time domain by
defining weights $w_i$ for each datum $Y_i$ measured at time $t_i$ and
then summing $w_i Y_i$ for all $i$.

This method of separating frequency space into broad bands has
the convenience of simplicity since convolving with Gaussians in the
time domain corresponds to multiplying by Gaussians in the Fourier
domain and the filter parameters are easy to determine. However, the
shape of the filter corresponding to the differences between levels is
awkward since it is skewed and it reduces the amplitude of the signal
everywhere except at one off-center frequency in the band. It is
preferable to have a filter that is flat over the band, at least in
the limit that errors $\sigma_i$ are equal for all $i$, with a
symmetric smooth drop-off at the lower and upper limit of the
band. One example of that is a filter defined by the following convolution~:
\begin{eqnarray}
H(\omega) &=& \int\limits_{-\infty}^{\infty} \textrm{d}\omega'
{1\over\delta_\omega\sqrt{\pi}}\exp\left[{-(\omega-\omega')^2
\over\delta_\omega^2}\right]
\Pi(\omega') \nonumber \\
\Pi(\omega) &=& 1\ \ \textrm{for}\ \ 
\vert\omega\vert < \Delta_\omega \nonumber\\
&=&0\ \ \textrm{for}\ \ \vert\omega\vert >\Delta_\omega
\label{bandfiltfreq}
\end{eqnarray}
Band-pass filtering, i.e. multiplying in the Fourier domain by this filter
function $H(\omega-\omega_c)$, centered on a frequency $\omega_c$,
corresponds in the time domain by convolving with the function~:
\begin{equation}
h(t) = {\sin\Delta_\omega t\over \pi t}
\exp\left[-{1\over 4}\delta_\omega^2 t^2\right]\cos\omega_c t
\label{bandfilttime}
\end{equation}
One can create successive bands that are orthogonal by choosing for 
instance~:
\begin{eqnarray}
{\delta_\omega\over\Delta_\omega} &\equiv& r\ll 1 \ \ 
\textrm{fixed}\nonumber\\
\Delta_{\omega,j+1}&=& \alpha_j\Delta_{\omega,j} \nonumber\\
{\omega_{c,j+1}\over\Delta_{\omega,j+1}} &\equiv& \gamma_{j+1}= 1 +
{\gamma_j+1\over\alpha_j}\ \ \textrm{with}\ \ \gamma_1 = 0
\label{filtpardefs}
\end{eqnarray}
The weights $w_{m,ij}$ are therefore in this case set as~:
\begin{equation}
w_{m,ij} = {\Delta_{\omega,j}\over W_j\pi} 
h_j(x_{m,ij})
\end{equation}
where $W_j$ is again a normalisation factor and the $x_{m,ij}$ and $h_j$ are~:
\begin{eqnarray}
x_{m,ij} &=& \Delta_{\omega,j} (t_m - t_i) \nonumber\\
h_j(x) &=& {\sin x\over x}\exp\left[-{1\over 4}r^2
x^2\right]\cos\left( \gamma_j x\right)
\label{xwdefs}
\end{eqnarray}
The band-pass filtered data $R$ are now calculated as~:
\begin{equation}
R_{i,j} \equiv R_{t_i,j} = \sum_m w_{m,ij} Y_m 
\label{broadfilter}
\end{equation}

\subsection{local oscillator step}
In order to obtain the frequency content of the time series, without
performing Fourier transforms on unevenly sampled data, one can employ 
what is known as a Hilbert transform. In radio receiver technology
this would be referred to as mixing the signal with a local
oscillator, and then applying a low-pass filter. The frequency $f$ of
the local oscillator is `tunable' : the width of the band-pass set by
the previous step is covered, sampling regularly in frequency with
some sampling step $\Delta f$. This sampling distance $\Delta f$ is 
chosen in combination with the width of the low-pass filter subsequently
applied, so that some oversampling is performed. 

The low pass filter to apply to the `mixed signal' is chosen to be a
Gaussian. This is a convenient choice because multiplying by a
(narrow) Gaussian filter in the frequency domain, corresponds to convolving
with a (wide) Gaussian in the time domain. The width $\Delta_{LPF}$ in 
the time domain of what corresponds to a low-pass filter, is linearly
related to the parameter $\Delta$ of the multiresolution level within which
the signal is to be frequency analysed.

One proceeds as follows. For each frequency $f_k$ define sets of
`local oscillator' weights 
$q_{i,k}$, $p_{i,k}$~:
\begin{eqnarray}
q_{i,k} &=& 2\cos(2\pi f_k t_i)\nonumber\\
p_{i,k} &=& -2\sin(2\pi f_k t_i)
\end{eqnarray}
and also define a set of low-pass filter weights $z_{i,l}$~:
\begin{equation}
z_{i,l} = {1\over\Delta_{LPF}\sqrt{\pi}}\exp\left[-\left({t_i -T_l\over
\Delta_{LPF}}\right)^2\right]
\label{lowpassfilter}
\end{equation}
The central times $T_l$ are spaced by some factor of order unity times 
$\Delta_{LPF}$. $\Delta_{LPF}$ is large in order to only let through
signal in a narrow band, and therefore the number of times $T_l$ is
small~: typically of the order of the number of nights of
observations. One can define a cosine-weighted average $\langle
R_{\cos}\rangle$ and a sine-weighted average $\langle R_{\sin}\rangle$
as follows~:
\begin{eqnarray}
\langle R_{\cos}\rangle_{lkj} = \sum_i z_{i,l}q_{i,k}R_{i,j}\nonumber\\
\langle R_{\sin}\rangle_{lkj} = \sum_i z_{i,l}p_{i,k}R_{i,j}
\label{Hilbertterms}
\end{eqnarray} 
The three indices for these two averaged quantities refer to~:
\begin{enumerate}
\item{} $j$ the `resolution level' of the time series, which
corresponds to a broad range in frequencies between a lower and upper
bound set by a smoothing width $\Delta_\omega$.
\item{} $k$ the central frequency $f_k$ of the tunable narrow-band filter runs
from the lower to the upper limit of the band $j$ to explore the
amplitude of signal in the time series at/around each $f_k$.
\item{} $l$ the central time on which a broad Gaussian is centered,
providing a narrow band filter. By taking a finite value for the width
of this Gaussian (rather than infinite) and stepping through the time
series, a time-frequency analysis is done, since it provides a
mechanism to consider the frequency content per night, or per few 
successive nights.
\end{enumerate}
For data covering only a few nights with a large gap during the
day-time, the window function has quite strong daily sidelobes which 
complicates analysis of a spectrum if there are a number of peaks
present, as is common in multi-periodic variable stars.
It is therefore useful to consider improvements on the method with the
aim of reducing sidelobe structure. This may be achieved by choosing
the low-pass filtering weights $z_{i,l}$ in a different manner than
described by (\ref{lowpassfilter}). 

\subsection{Strategy}
Consider again the two summations of Eq. (\ref{Hilbertterms})
which are defined to be functions of frequency $f$. The weights are
now $\zeta_i$ and in principle one wishes to be free to choose weights
differently depending on the frequency $f_k$ of the local oscillator.
The time $T_l$ on which the low-pass filter is centered in Eq.
(\ref{lowpassfilter}) now enters the discussion by explicitly setting
the reference time for the  phase $\phi$,
which means replacing $t_i$ with $t_i-T_l$. With the weights $\zeta$
yet to be determined two functions of $f$ are defined as follows~:
\begin{eqnarray}
\sum_i \zeta_{i,kl}\cos\left(2\pi (f-f_k) (t_i -T_l) \right) &\equiv&\Xi (f) 
\nonumber\\
\sum_i \zeta_{i,kl}\sin\left(2\pi (f-f_k) (t_i -T_l) \right) &\equiv&X (f)
\label{KsiXdefs}
\end{eqnarray}
The method has three separate aims to achieve with its choice for the 
weights. These aims may well not be perfectly compatible and therefore
require a form of compromise~:
\begin{itemize}
\item{} The first function $\Xi(f)$ needs to be a function peaked at
$f=0$ and smoothly dropping away, with as little side-lobe structure
as possible. For instance~:
\begin{equation}
\Xi(f) \approx {1\over\sqrt{\pi}\Delta_f}\exp\left[-\left({f-f_k\over\Delta_f}
\right)^2\right]
\label{ksitarget} 
\end{equation}
with as small a width $\Delta_f$ as achievable.
\item{} The second function $X(f)$ needs to be as close to $0$ as 
possible everywhere
\item{} The errors in the resulting amplitudes $A^2$ need to be kept
as small as possible.
\end{itemize}

\begin{figure}
\centerline{\psfig{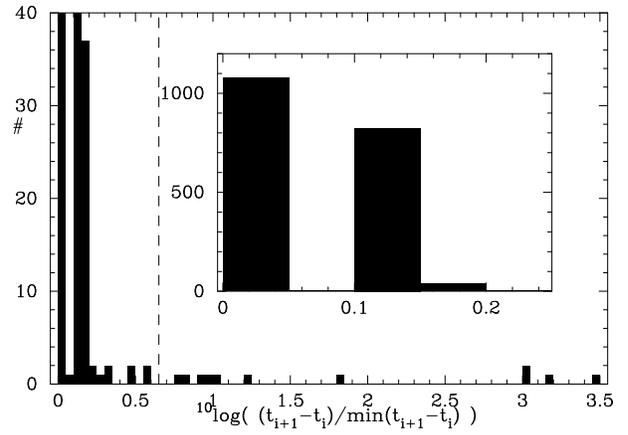}} 
\caption[]{A histogram of the distribution of intervals between successive 
sampling times for an artificial time series, divided by the smallest of 
these intervals. The vertical dashed line indicates the mean time interval
for the entire set. The inset shows the part of the distribution at small 
intervals which is off-scale for the larger figure.}
\label{Thistfig}
\end{figure}

In the framework of the SOLA method (cf. \cite{PT92}, \cite{PT94}) 
the best compromise between these three aims
can be achieved by minimising for $\zeta_{i,k}$ the quantity ${\cal
W}(\zeta)$ defined by~:
\begin{eqnarray}
{\cal W}(\zeta) \equiv (1-\mu)\int\limits_{f_{\rm low}}^{f_{\rm
high}} {\rm d}f \left[\Xi(f) - {\cal T}(f)\right]^2  \nonumber\\
+\mu \int\limits_{f_{\rm low}}^{f_{\rm high}}{\rm d}f 
\left[X(f)\right]^2
+ \lambda\sum_{i,j} \zeta_{i,k}\zeta_{j,k}\Sigma_{ij}
\label{SOLAmin}
\end{eqnarray}
while taking into account the normalisation constraint~:
\begin{equation}
\sum_i \zeta_{i,k}\left[\int\limits_{f_{\rm low}}^{f_{\rm
high}} {\rm d}f \cos\left(2\pi (f-f_k) (t_i -T_l) \right)\right] \equiv 1
\end{equation}
Here the target function ${\cal T}(f)$ is taken to be a Gaussian as
in (\ref{ksitarget}). The `target function' for $X(f)$ is $0$. The
matrix $\Sigma_{ij}$ is the variance-covariance matrix of the data
errors. It is usual to assume that the errors on the data are Gaussian 
distributed and independent, so that $\Sigma$ is a diagonal matrix,
but this is not essential for the algorithm. The methods described in
\cite{Ko99} to estimate or model error properties from residuals
using ARIMA models can be applied here as well. The parameters
$\mu\in[0,1]$ and $\lambda\in[0,\infty)$ weight the relative importance of the
three terms. Their value needs to be set by trial and error. A very
low choice of $\lambda$ normally leads to unacceptably high errors on
the inferred $A^2$, whereas a very high value generally produces the
undesirable effect of increasing width or structure in $\Xi$. Similar
arguments hold for $\mu$.

\begin{figure*}
\centerline{\psfig{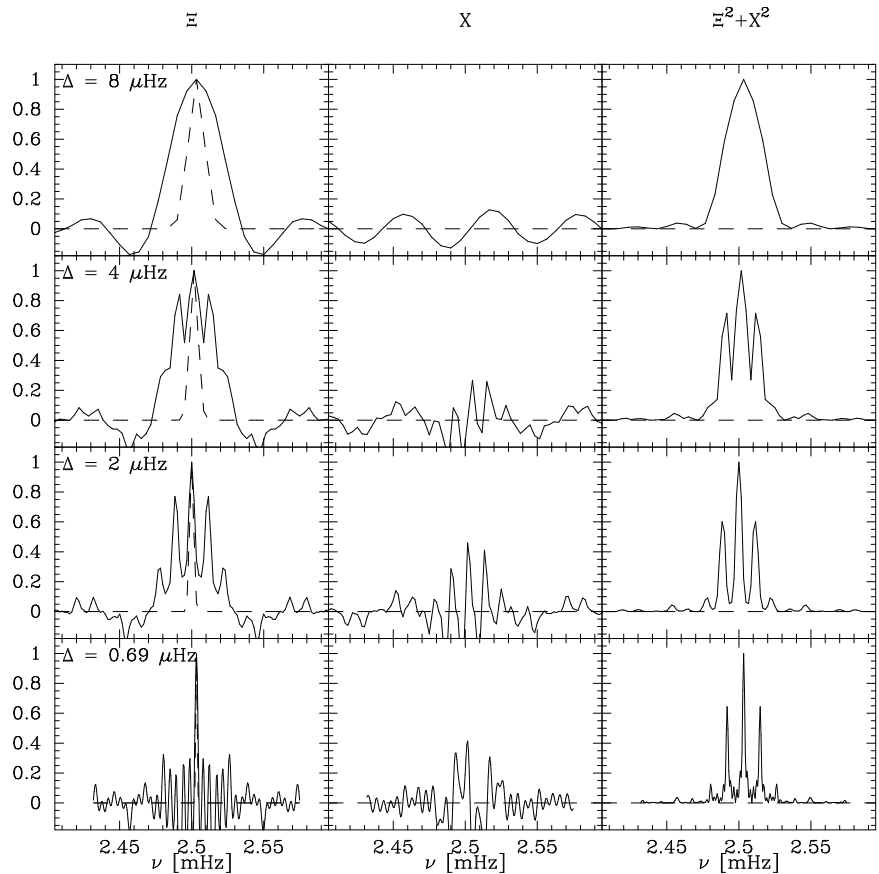}} 
\caption[]{The response to a monochromatic signal of $2.5\ {\rm mHz}$ sampled
with the sampling shown in Fig. 1, which is the equivalent of the window
function for this method. The left and middle columns show the
functions $\Xi$ and $X$, with in addition the target functions (dashed
linmes). The right hand column shows the sum of the
square of these which is the equivalent of the window function for
power. The rows are in order of increasing spectral resolution with
the $\Delta_f = 8,\ 4,\ 2,\ 0.69\ \mu{\rm Hz}$, from top to bottom.
The lower and upper frequency limit of the band are set to $f_{\rm
low} = 2.5\ {\rm mHz}$ and $f_{\rm high} = 6\ {\rm mHz}$.}
\label{tableauFig}
\end{figure*}

A point that {\bf is} crucial for this version of the SOLA algorithm,
is that $f_{\rm low}$ and $f_{\rm high}$ must be finite. This is
necessary because the integrations in Eq. (\ref{SOLAmin}) are
over products of cosines, which causes problems if they are to be
integrated over an infinite range. The algorithm is therefore limited
in application to bandwidth-limited data. In practice this is not a
severe problem, since by construction of the multiresolution steps, 
bandwidth limitation is achieved. Also, given a finite minimum
interval between measurements it is known that signals with
frequencies above the Nyquist frequency $\propto 1/\min(t_{i+1}-t_i)$ 
cannot be resolved by such data, which also sets an upper limit to
the frequency.

\section{Results for the sampling of an asteroseismic campaign}
\begin{figure}
\centerline{\psfig{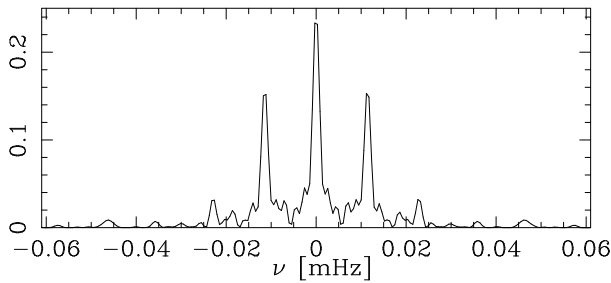}} 
\caption[]{The window function obtained using the algorithm of section
2.1 with parameter choices corresponding to the bottom right panel of
Fig. 2.}
\label{GaussHilbert}
\end{figure}
To show how the methods work, the sampling times are used of a time 
series obtained during an asteroseismic campaign~: the 
sampling times are nearly regularly spaced during blocks of time, with
large gaps in between the blocks. This corresponds to night-time observing 
for a few successive nights, with day-time interruptions as well as loss 
of data due to adverse weather conditions as well as some technical glitches. 
The total number of sampling times is $1962$, the median sampling rate
is about $\sim 70\ s$ whereas the mean rate is x$\sim 236\ s$.
\begin{figure*}
\centerline{\psfig{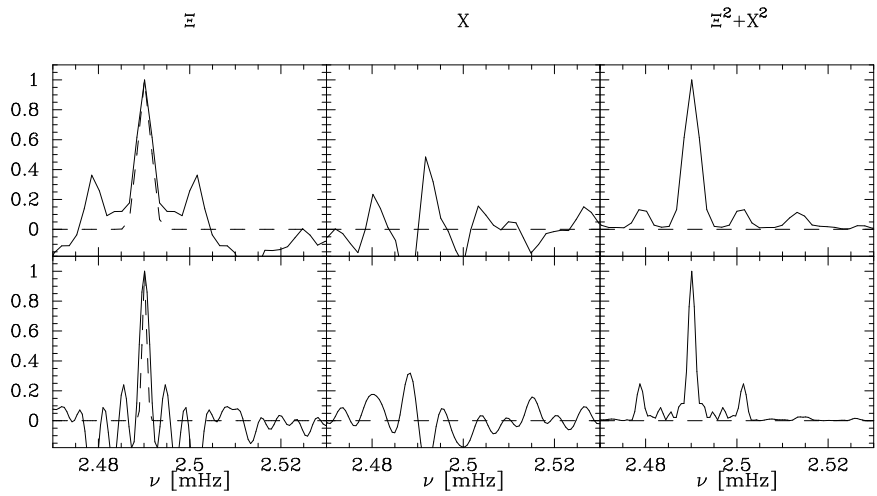}} 
\caption[]{The effect of reducing the width of the broad band over
which the frequencies are sampled on the window function. Here the
lower and upper frequency limit of the band are set to $f_{\rm low} =
2.456\ {\rm mHz}$ and  $f_{\rm high} = 2.544\ {\rm mHz}$. The target
sampling frequency is $2.49\ {\rm mHz}$.}
\label{midreswin}
\end{figure*}
If the series had been regularly sampled at the median rate the
Nyquist frequency would have been $\sim 7.14\ {\rm mHz}$.
Fig. \ref{Thistfig} shows the distribution of sampling time intervals
between successive measurements, clearly demonstrating that the time
series is not fully regularly sampled. Using these sampling times the
linear coefficients $\zeta$ are determined and then used on an
artificial signal with a frequency close to $2.5\ {\rm mHz}$ to obtain
a window function, shown in Fig. \ref{tableauFig}. Also shown for
comparison in Fig. \ref{GaussHilbert} is the window function obtained
from the method of Sect. 2.1. 
The rows in Fig. \ref{tableauFig} are in order of increasing spectral
resolution with the $\Delta_f = 8,\ 4,\ 2,\ 0.69\ \mu{\rm Hz}$, from
top to bottom, where $0.69 \mu{\rm Hz}$ corresponds to the resolution
limit set by the full length of the time series if it were regularly
sampled over this period. 

Perhaps disappointingly the spectral resolution that can be attained
over the band is quite poor. However, the quality of the window function 
depends on the ratio of the resolution over the width of the band~: $\Delta_f
/\Delta_\omega$. By bandwidth filtering of the data in the same way as
discussed in section 2, but over a narrower band, a better resolution
can be achieved within the band. In Fig. \ref{midreswin} the
resolution $\Delta_f$ is set to $2\ \mu{\rm Hz}$ and the limits of the
band are $f_{\rm low} = 2.456\ {\rm mHz}$ and $f_{\rm high} = 2.544\
{\rm mHz}$ i.e. a factor of $40$ narrower than before. The target sampling
frequency in this case is $2.49\ {\rm mHz}$. Now the FWHM of
the window function in power is $3.3\ \mu{\rm Hz}$ (top right panel of
Fig \ref{midreswin}). Even at a target width $\Delta_f = 0.69\ \mu{\rm
Hz}$ the sidelobe structure in the window function for power  
(bottom right panel of Fig \ref{midreswin}) is much reduced compared
to the same resolution for the previous case (the bottom right panel
of Fig. \ref{tableauFig}).

\section{Discussion}
The method described in this paper, for the analysis of multiperiodic,
irregularly sampled time series, has several potential uses. The first
point to note is that the method is a time-frequency method formulated
in the time domain. This means that the frequencies of the signals in an
irregularly sampled time series can be monitored as a function of
epoch. This allows for the possibility to detect changes in
oscillation frequencies with epoch. Such changes could arise in
asteroseismology through similar mechanisms that produce changes in
oscillation frequencies of the Sun over the solar cycle and are
therefore of interest to determine. 

The second point to note is that the method is linear. The algorithm
produces a set of linear coefficients on the basis of the sampling
times alone. These coefficients can then trivially be combined with
data, and with measurement errors, in order to provide the power or
amplitude of the time series and also an error estimate for this
power, and even a phase if required. Evidently this linearity is
convenient for standard asteroseismic data acquisition which provides
photometry or velocity. However, it is arguably even more useful if
the data obtained is more complex such as a time series of images. A
particular application lies in addressing another fundamental problem
in asteroseismology, which is the identification of the surface node
line pattern associated with any particular frequency~: the mode
identification problem. For stars with a single dominant mode, long
baseline optical interferometry has been suggested as a means to image
the flux variation over the surface, thus allowing the identification
of the mode (cf. \cite{Jan+01} for a description of the relevant
technique). In particular the closure phase of interferometric
signals between three telescopes is sensitive to symmetries in the
flux distribution over the surface of stars (cf. \cite{DdS+05}). 
For multiperiodic stars
such a technique would suffer since the superposition of patterns
would reduce the closure phase response. However, the acquisition of
interferometric fringes is already done repeatedly as a function of
time for operational reasons. By combining closure phase signals at
the appropriate phase in time one could select a single frequency to
image, thus restoring the closure phase signal strength. This
technique might be referred to as `stroboscopic interferometry'. The
closure phase is a somewhat more complex form of data than time series
of velocities or fluxes, and carrying out stroboscopic interferometry in
practice is facilitated by being able to carry out the time-filtering
using the coefficients generated with this SOLA based algorithm.
Appropriate tests of the algorithm for this application are in
progress (Pijpers, in preparation).

A third point to note is that this method can also be adapted to
streamline procedures for model fitting in asteroseismology.
The Gaussian form of the target function chosen here is
intended to obtain measures of the oscillation power or amplitude as a
function of frequency. However, in fitting models to a given star it
is not just a single frequency but the entire pattern of frequencies
for which a best match needs to be found. This can be achieved more
directly by calculating the expected oscillation spectrum for each
model and taking for the target function ${\cal T}(f)$ not a single
Gaussian, but a sum of Gaussians centered on model frequencies
$f_{nl}$, in which the $nl$ are the radial mode order $n$ and azimuthal
degree $l$, instead of the single $f_k$. The coefficients obtained in
this way, when combined with the data, will produce a maximal response
for that model which has a set of frequencies that is the closest match
to the data. Further tests of this are necessary to establish the
error propagation properties of such an approach, including the
influence of systematic effects (Pijpers, in preparation).
\section*{Acknowledgments}
The method described in section 2.2 formalises, within the framework of
the SOLA algorithm (\cite{PT92}, 1994), an idea originally proposed by
H. Kjeldsen (private communication), who is thanked for suggesting to
the author to pursue this.


\end{document}